\documentclass{ws-ijgmmp}

\usepackage{xcolor}
\usepackage[verbose,hypertexnames=false]{hyperref}
\hypersetup{colorlinks=false,allbordercolors=blue,pdfborderstyle={/S/U/W 1}}

\usepackage[a4paper,top=3.5cm,bottom=3cm,left=2.5cm,right=2.5cm]{geometry}
\usepackage{amssymb}
\usepackage{amsmath}
\usepackage{physics}
\usepackage{mathtools}
\usepackage{tipa}
\usepackage{latexsym} 
\usepackage{graphicx}
\usepackage{slashed}
\usepackage{bbold}
\usepackage{cancel}
\usepackage{comment}
\usepackage{color}
\usepackage{graphicx,epsfig,color}
\usepackage[dvipsnames]{xcolor}
\usepackage{comment}
\usepackage{cite}
\usepackage{tikz}
\usepackage[compat=1.1.0]{tikz-feynman}
\usepackage{caption}
\usepackage{subcaption}

\newcommand{\be}{\begin{equation}}
	\newcommand{\ee}{\end{equation}}
\newcommand{\bea}{\begin{eqnarray}}
	\newcommand{\eea}{\end{eqnarray}}
\newcommand{\vv}{``}

\newcommand{\mpl}{M_P}

\begin{document}

\markboth{C. Branchina, V. Branchina, F. Contino, R. Gandolfo and A. Pernace}
{Gravity and the Higgs boson mass}

%
\catchline{}{}{}{}{}
%

\title{Gravity and the Higgs boson mass}

\author{C. Branchina$^{1\,a\,b}$,
	V. Branchina\footnote{Corresponding author}\,\,\,$^{\,c\,d}$\,\,\,, 
	F.
	Contino$^{2\,e,\,f}$,
	R. Gandolfo$^{3\,c\,d}$,
	A.
	Pernace$^{4\,c}$}

\address{
	${}^a$
	Department of Physics, University of Calabria,
	Arcavacata di Rende, I-87036, Cosenza, Italy\\
	${}^b$
	INFN-Cosenza,
	Arcavacata di Rende, I-87036, Cosenza, Italy\\
	${}^c$Department of Physics, University of Catania, Via Santa Sofia 64, I-95123 
	Catania, Italy\\
	${}^d$INFN-Catania,
	Via Santa Sofia 64, I-95123 
	Catania, Italy\\
	${}^e$
	Scuola Superiore Meridionale, Largo San Marcellino 10, 80138 Napoli, Italy\\
	${}^f$
	INFN-Napoli, Complesso Universitario di Monte S. Angelo, Via Cinthia Edificio 6, 80126 Napoli, Italy\\
	\vskip 6pt
	\email{${}^1$carlo.branchina@unical.it}
	\email{${}^*$vincenzo.branchina@ct.infn.it} 
	\email{${}^2$f.contino@ssmeridionale.it}
	\email{${}^3$riccardo.gandolfo@ct.infn.it}
	\email{${}^4$arcangelo.pernace@ct.infn.it} 
}

\maketitle

\begin{history}
\received{(Day Month Year)}
\revised{(Day Month Year)}
\accepted{(Day Month Year)}
\published{(Day Month Year)}
\end{history}

\begin{abstract}
According to usual calculations, both in flat and curved spacetime, the mass $m^2$ of a scalar particle is quadratically sensitive to the ultimate scale of the theory, the UV physical cutoff $\Lambda$.
Building on previous work \cite{Branchina:2024lai,Branchina:2024xzh,Branchina:2025lqw}, here we calculate the one-loop effective action $\Gamma^{1l}$ for a scalar field on a spherical gravitational background, using the diffeomorphism invariant Fradkin-Vilkovisky path integral measure. This measure gives rise to an automatically dimensionless fluctuation determinant, that we calculate resorting to a numerical cut $N$ on the number of eigenvalues. We show that the UV-sensitivity of the radiative correction $\delta m^2$ to $m^2$ depends crucially on the relation between $N$ and the UV cutoff $\Lambda$. We find that if the transition from $N$ to $\Lambda$ is realized through the off-shell background radius, $\delta m^2$ presents the well-known quadratic sensitivity to $\Lambda$. If, instead, the transition is realized resorting to the on-shell radius that minimizes the classical action, the mass turns out to be only logarithmically sensitive to $\Lambda$.
\end{abstract}

\keywords{one-loop effective action; scalar mass; naturalness problem.}

\ccode{Mathematics Subject Classification 2020: 81T20, 81T15, 83C47}

\section{Introduction}
\label{Introduction}
\numberwithin{equation}{section}

The Standard Model (SM) of particle physics is one of the greatest achievements of modern theoretical and experimental physics, a synthesis started in the late sixthies/early  seventhies of the last century with the emergence of the electroweak and strong interaction theory. The SM, that has received several experimental confirmations over the years, among which the discovery of the Higgs boson at LHC \cite{ATLAS:2012yve,CMS:2012qbp}, provides the basic ground to describe and understand a great variety of phenomena. Still, it is not a fully fledged theory. It is an effective field theory (EFT) valid up to a maximal energy scale, the UV physical cutoff, that we generically indicate with $\Lambda$ (say GUT scale $M_{\rm GUT}$, Planck scale $\mpl$, string scale $M_s$). 

Typical approaches to go beyond the SM consider supersymmetric extensions and/or models where the Higgs boson appears as a composite particle. In both cases, the UV completion is a field theory whose validity extends to energy regimes higher than those typically explored in SM physics. One of the motivations for supersymmetry is to realize the cancellation (thanks to the presence of superpartners) of quadratic divergent contributions to the Higgs mass. 
In fact, performing the calculation of the propagator and/or the effective action both in flat and curved spacetime, the bare Higgs boson mass $m^2(\Lambda)$ receives contributions $\delta m^2$ proportional to  $\Lambda^2$, due to unsuppressed quantum fluctuations (no symmetry protection). 
In this respect, it is important to stress that, if regularization schemes as dimensional or zeta function regularization are used, this quadratic UV sensitivity of the mass is not seen. However, these regularizations operate by construction a cancellation of power-law divergences and cannot be regarded as a physical mechanism responsible for the suppression of the aforementioned strong UV sensitivity \cite{Branchina:2022jqc}.

As stressed above, the SM has to be considered as an EFT valid up to the UV scale $\Lambda$. The fact that $\delta m^2 \sim \Lambda^2$ means that $m^2(\Lambda)$ must be $\mathcal O(\Lambda^2)$ too, with a coefficient that has to be finely tuned for the Higgs boson mass $m^2_{\text{\tiny H}}=m^2(\Lambda)+\delta m^2$ to coincide with the measured value $\sim(125\, \text{GeV})^2$. Since $\Lambda\gg m_{\text{\tiny H}}$, 
the bare mass $m^2(\Lambda)$ has to be taken {\it unnaturally} large\footnote{For recent discussions on naturalness, renormalization and the flow of the mass see \cite{Hariharakrishnan:2024iba,Yamada:2020bqe,Aoki:2012xs,Gies:2025pqv,Branchina:2022gll,Branchina:2022jqc,Marian:2024nfi,Garces:2025rgn,Wells:2025hur,Peskin:2025lsg}.} with respect to $m^2_{\text{\tiny H}}$  \cite{Georgi:1974yf,tHooft:1979rat,Veltman:1980mj,Bardeen:1995kv,Giudice:2013yca}. 
This is one aspect of the so-called naturalness problem for the Higgs boson mass, to which we will refer in the following as \vv physical cutoff problem'' (PCP).

Another aspect of this long-standing issue arises when the SM is viewed as embedded in a higher energy theory where the Higgs field $H(x)$ is coupled to fields of large masses. Let us consider for instance 
a supersymmetric extension of the SM,
where SUSY is broken by 
a large stop mass $\widetilde m_t \gg m_{\text{\tiny H}}$. The Higgs mass receives the correction ($y_t$ is the top Yukawa coupling, $\mu$ the renormalization/subtraction scale)
\be\label{stop}
\delta m^2 \sim y_{t}\, \widetilde m_t^2\, {\rm ln}\frac {\widetilde m_t^2}{\mu^2}\,,
\ee 
and again we have to cope with a quadratic radiative correction to the Higgs boson mass. In the following, we refer to this other aspect of the naturalness problem as \vv large masses problem'' (LMP).

Building on previous work \cite{Fradkin:1973wke, Fradkin:1975sj, Unz:1985wq, Branchina:2024lai, Branchina:2024xzh, Branchina:2025kmd, Branchina:2025lqw}, in this paper we revisit the PCP. We calculate the one-loop effective action $\Gamma^{1l}$ for a scalar field on a spherical gravitational background of radius $a$. As shown in \cite{Fradkin:1973wke, Fradkin:1975sj}, and later in \cite{Branchina:2025lqw}, the diffeomorphism invariant path integral measure to be used in this calculation is the Fradkin-Vilkovisky (FV) measure. This measure leads to an automatically dimensionless fluctuation operator, and this allows to calculate the determinant introducing a numerical cut $N$ on the number of eigenvalues. Crucial is the relation between $N$ and the UV cutoff $\Lambda$. We show that if such a relation is realized resorting to off-shell background radius $a$,
\begin{equation}\label{offshell}
	\Lambda=\frac{N}{a}\,,
\end{equation}
the radiative correction $\delta m^2$ presents the well-known quadratic sensitivity to $\Lambda$. However, another choice is possible when considering a spherical gravitational background. We can resort to the on-shell radius $a_{\rm m}$ that minimizes the classical action, 
\begin{equation}\label{onshell}
	\Lambda=\frac{N}{a_{\rm m}}\,.
\end{equation}
We find that using this latter relation leads to a result for $\delta m^2$ that depends only logarithmically on $\Lambda$. Therefore, the embedding of the theory in a non-trivial gravitational background seems to open to the possibility of a mass with only a mild dependence (logarithmic) on the UV scale $\Lambda$, thus dispensing with the necessity of fine-tuning.

For the purposes of our analysis, it is not necessary to take into account the full Higgs sector of the SM; it is sufficient to consider the case of a single component scalar field on a non-trivial gravitational background. We then calculate and analyse the one-loop effective action $\Gamma^{1l}$ for this simpler theory. 

The rest of the paper is organised as follows. In section \ref{1lEA}, we calculate the one-loop effective action $\Gamma^{1l}$ for a scalar field non-minimally coupled to gravity on a spherical gravitational background. In section \ref{1lparameters}, we derive the one-loop corrections to the parameters of the theory (boson mass, non-minimal gravitational coupling, self-interaction coupling) considering the two relations between $N$ and $\Lambda$ mentioned above. Section \ref{conclusions} is for the conclusions.

\section{One-loop effective action}
\label{1lEA}
Let us add to the purely gravitational Einstein-Hilbert (Euclidean) action the contribution of a single component real scalar field $\phi$ non-minimally coupled to gravity

\begin{equation}\label{claction}
	S[g_{\mu\nu},\phi]=\frac{1}{16\pi G}\int\dd[4]x\,\sqrt{g}\,\left(-R+2\Lambda_{\rm cc}\right)+\int\dd[4]x\,\sqrt{g}\left[\frac12\,g^{\mu\nu} \partial_\mu\phi\,\partial_\nu\phi+\frac{\xi}{2}R\phi^2+V(\phi)\right]\,.
\end{equation}
Taking for $g_{\mu\nu}$ the metric $g^{(a)}_{\mu\nu}$ of a sphere of radius $a$, the action\,\eqref{claction} becomes ({\footnotesize $\int\dd[4]x\,\sqrt{g^{(a)}}=\frac{8\pi^2}{3}a^4$\,,\, $R(g^{(a)})=\frac{12}{a^2}$})
\begin{equation}
	S^{(a)}[\phi]=\frac{\pi\Lambda_{\rm cc}}{3G}a^4-\frac{2\pi}{G}a^2+\int\dd[4]x\,\sqrt{g^{(a)}}\left[\frac12\,g^{(a)\,\mu\nu} \partial_\mu\phi\,\partial_\nu\phi+\frac{\xi}{2}\,\frac{12\,}{a^2}\,\phi^2+V(\phi)\right]\,.
	\label{scalaraction2}
\end{equation}
Different powers of $a$ correspond to different gravitational terms. We will use this feature to identify the gravitational operators in the one-loop correction $\delta S^{1l}$ to $S^{(a)}$. To calculate $\delta S^{1l}$, we resort to the background field method\,\cite{Abbott:1980hw,Abbott:1981ke} and write $\phi=\Phi+\eta$, where $\Phi$ is a constant background and $\eta$ the fluctuation. Expanding $S^{(a)}[\phi]$ around $\Phi$ up to quadratic terms in $\eta$, we have
\begin{align}\label{effac1}
	e^{-\delta S^{1l}}=\int \big[\mathcal{D}u(\eta)\big]\, e^{-S_2}\,,
\end{align}
where
\be\label{S2_1}
S_2\equiv\frac12\int\dd[4]x\,\sqrt{g^{(a)}}\,\,\eta\Big[-\square_{\,a}+\frac{12\,\xi}{a^2}+V''(\Phi)\Big]\eta\,,
\ee
with $-\square_{\,a}$ the spin-$0$ Laplace-Beltrami operator for a sphere of radius $a$ and $V''(\Phi)$ the second derivative of the potential with respect to $\Phi$. The measure $\big[\mathcal{D}u(\eta)\big]$ is given by
\be \label{meas}
\big[\mathcal{D}u(\eta)\big]=\prod_x\left[\left(g^{(a)\,00}(x)\right)^{\frac12}\left(g^{(a)}(x)\right)^{\frac14}\dd{\eta(x)}\right]\,,
\ee
where the factors $\left(g^{(a)\,00}(x)\right)^{\frac12}\left(g^{(a)}(x)\right)^{\frac14}$ arise from the integration over the conjugate momenta of $\phi(x)$ in the original (Hamiltonian) formulation of the theory \cite{Fradkin:1973wke, Fradkin:1975sj,Unz:1985wq,Branchina:2024lai,Branchina:2024xzh, Branchina:2025kmd, Branchina:2025lqw}. 
It is worth to stress here that, despite the presence of $g^{(a)\,00}$ factors, this measure is
diffeomorphism invariant. This invariance emerges from a delicate balance between the different elements involved in the definition of
the path integral. Among them, the necessity of
introducing a time ordering parameter and a discretization (lattice) of spacetime. Under a general coordinate transformation, the time ordering parameter and the lattice both transform, and this induces the appearance of non-trivial terms. The $g^{(a)\,00}$ factors in\,\eqref{meas} ensure the cancellation of these non-trivial terms, and ultimately guarantee the
diffeomorphism invariance\footnote{In\cite{Bonanno:2025xdg} a different  conclusion is reached. As shown in \cite{Branchina:2025lqw}, the reason for such a difference is that the authors of \cite{Bonanno:2025xdg} miss the non-trivial terms mentioned above.} of the measure \cite{Fradkin:1973wke, Branchina:2025lqw}.
From the invariance of both $\big[\mathcal{D}u(\eta)\big]$ in\,\eqref{meas} and $S_2$ in\,\eqref{S2_1}, we have that $\delta S^{1l}$ in\,\eqref{effac1} is also invariant. 

For our purposes, it is convenient to calculate $\delta S^{1l}$ considering coordinate systems (as for instance the four angles that parametrize the sphere) where the metric $g^{(a)}_{\mu\nu}$ can be written as 
\begin{equation}\label{metric1}
	g^{(a)}_{\mu\nu}=a^2 \,\widetilde g_{\mu\nu}\,,
\end{equation}
where the elements of\, $\widetilde g_{\mu\nu}$ are dimensionless and $a$-independent. Using\,\eqref{metric1}, the factor {\small $\left(g^{(a)\,00}(x)\right)^{\frac12}$ $\left(g^{(a)}(x)\right)^{\frac14}$} in\,\eqref{meas} is written as
\begin{equation}\label{measterms}
	\left(g^{(a)\,00}(x)\right)^{\frac12}\left(g^{(a)}(x)\right)^{\frac14}=a\left(\widetilde g^{\,00}(x)\right)^{\frac12}\left(\widetilde g(x)\right)^{\frac14}\,.
\end{equation}
As we will see, the effect of these terms in the measure\,\eqref{meas} is conveniently taken into account if we define the dimensionless field
\begin{equation}\label{hateta}
	\widehat \eta \equiv a\eta\,.
\end{equation}
Inserting\,\eqref{metric1} and\,\eqref{hateta} in\,\eqref{S2_1}, for $S_2$ we have
\begin{equation}\label{S2_2}
	S_2=\frac12\int\dd[4]x\,\sqrt{\widetilde g}\,\,\widehat\eta\Big[-\widetilde \square+12\,\xi+ a^2 \,V''(\Phi)\Big]\widehat\eta\,,
\end{equation}

\noindent
where $-\widetilde\square$ is the dimensionless spin-$0$ Laplace-Beltrami operator  defined as
\begin{equation}\label{LB1}
	-\widetilde\square\equiv-a^2\,\square_{\,a}\,.
\end{equation}
We now observe that, since
\begin{equation}\label{prodh}
	\dd{\eta(x)}=a^{-1}\dd{\widehat \eta(x)}\,,
\end{equation}
after insertion of\,\eqref{measterms} and\,\eqref{prodh} in\,\eqref{meas}, the measure $\big[\mathcal{D}u(\eta)\big]$ becomes
\begin{equation}\label{meas2}
	\big[\mathcal{D}u(\eta)\big]=\Big[\prod_x \left(\widetilde g^{\,00}(x)\right)^{\frac12}\Big]\Big[\prod_x \left(\widetilde g(x)\right)^{\frac14}\dd{\widehat\eta(x)}\Big]\,.
\end{equation}
The factor $a^{-1}$ in\,\eqref{prodh} has been exactly compensated by the factor $a$ in\,\eqref{measterms}, eventually resulting in the $a$-independent path integral measure above.

Inserting\,\eqref{S2_2} and\,\eqref{meas2} in\,\eqref{effac1} we obtain
\begin{align}
	e^{-\delta S^{1l}}=\Big[\prod_x \left(\widetilde g^{\,00}(x)\right)^{\frac12}\Big]\int\Big[\prod_x \left(\widetilde g(x)\right)^{\frac14}\dd{\widehat\eta(x)}\Big]\, e^{-\frac12\int\dd[4]x\,\sqrt{\widetilde g}\,\,\widehat\eta\big[-\widetilde \square+12\,\xi+ a^2 \,V''(\Phi)\big]\widehat\eta}\,.
	\label{oneloopeffac2}
\end{align}
Finally, performing the Gaussian integrations we get\footnote{Had we missed in the measure\,\eqref{meas} the factors { $\left(g^{(a)\,00}(x)\right)^{1/2}\left(g^{(a)}(x)\right)^{1/4}$}, the $a$-dependence of the fluctuation operator in\,\eqref{scalarres} would have been altered.} 
\begin{equation}\label{scalarres}
	\Gamma^{1l}=S^{(a)}[\Phi]+\frac12\Tr\log\left(-\widetilde\square+ 12 \xi + a^2V''(\Phi)\right)+\,\mathcal{C}\,,
\end{equation}
where ({\small $\delta^{(4)}(0)$} below is due to the usual replacement $\sum_x\to\int\text{\small$\dd[4]{x}$}$)
\begin{equation}\label{C}
	\mathcal{C}\equiv-\frac12\log\Big(\prod_x \widetilde g^{\,00}(x)\Big)=-\frac{\delta^{(4)}(0)}{2}\int\dd[4]{x}\log(\widetilde g^{\,00}(x))
\end{equation}
comes from the exponentiation of the measure term {\small $\prod_x (\widetilde g^{\,00}(x))^{1/2}$} (see\,\eqref{oneloopeffac2}). The presence of the non invariant term $\mathcal C$ might lead one to suspect that the above result for $\Gamma^{1l}$ is not invariant. This is not the case. In fact, as thoroughly discussed in \cite{Fradkin:1973wke, Fradkin:1976xa,Barvinsky:1985an}, and more recently in \cite{Branchina:2025lqw}, subtleties arise in the calculation of $\log \big(-\widetilde\square+ 12 \xi + a^2V''(\Phi)\big)$: one has to carefully take into account the distributional nature of the Green's function of the operator $(-\widetilde\square+ 12 \xi + a^2V''(\Phi))$ \cite{Fradkin:1976xa}. When this is done, from the calculation of $\Tr\log \big(-\widetilde\square+ 12 \xi + a^2V''(\Phi)\big)$ the non-trivial term {\small $\frac{\delta^{(4)}(0)}{2}\int\dd[4]{x}\log(\widetilde g^{\,00}(x))$} arises, that eventually cancels $\mathcal C$ in\,\eqref{scalarres}. This is why in the forthcoming expressions $\mathcal{C}$ does not appear. All the other terms coming from the calculation of \vv\,$\Tr\log$\,'' are diffeomorphism invariant \cite{Fradkin:1976xa}.

We have to calculate now the right hand side of\,\eqref{scalarres}. For the regularization of the trace, we will follow two different strategies: (i) we consider the sum over a finite number $N$ of eigenvalues; (ii) we repeat the calculation using proper-time regularization. Taking for the potential the truncation $V(\Phi)=\frac{m^2}{2}\Phi^2+\frac{\lambda}{4!}\Phi^4$, we will see that the radiative correction $\delta m^2$ to $m^2$ strongly depends on the way the numerical cut $N$ is related to the UV cutoff $\Lambda$. More specifically, if the transition from $N$ to $\Lambda$ is realized through the relation\,\eqref{offshell}, namely $N=\Lambda\,a$, we find that the usual quadratic divergence is present in $\delta m^2$. If, instead, the transition is realized resorting to the relation\,\eqref{onshell}, $\delta m^2$ shows only a logarithmic sensitivity to the UV scale $\Lambda$. We then see that considering a curved gravitational background seems to open to the possibility of having a mass that has only a mild dependence on the UV scale $\Lambda$. 

Let us begin with the calculation of the trace in\,\eqref{scalarres} considering the sum over the eigenvalues of the fluctuation operator. To this end, we recall that the eigenvalues $\lambda_n$ of the dimensionless Laplace-Beltrami operator $-\widetilde\square$ and the corresponding degeneracies $D_n$ are
\begin{equation}
	\lambda_n=n^2+3n\qquad ;\qquad D_n=\frac{1}{3}\left(n+\frac{3}{2}\right)^3-\frac{1}{12}\left(n+\frac{3}{2}\right)\,.
	\label{eigenvalues}
\end{equation}
The regularization is implemented taking a finite number $N$ ($\gg1$) of eigenvalues $\lambda_n$ ($n=0,1,\dots,N$). 
From\,\eqref{scalarres} we have (the choice of $N-2$ rather than $N$ as upper limit is for convenience)
\begin{align}
	\delta S^{1l}=\frac12\sum_{n=0}^{N-2} \Bigl[D_n\log\left(\lambda_n+ 12 \xi + a^2V''(\Phi)\right)\Bigr]\,.
	\label{calculation1}
\end{align}
The numerical UV cut $N$ in\,\eqref{calculation1} implements a gauge invariant regularization\footnote{The gauge invariance is guaranteed by the fact that $N$ is a cut on the eigenvalues $\lambda_n$.}. The connection between $N$ and the UV physical cutoff $\Lambda$ \cite{Branchina:2024lai,Branchina:2024xzh, Branchina:2025kmd} will be investigated in the next section.

Inserting\,\eqref{eigenvalues} in\,\eqref{calculation1}, and considering the identity {\small $\log\left(x/y\right)=-\int_{0}^{+\infty}\dd{u}\left[\left(x+u\right)^{-1}-\left(y+u\right)^{-1}\right]$}, the sum can be performed and put in closed form. For our purposes, it is sufficient to consider the expansion for $N\gg 1$. We get
\begin{align}
	\delta S^{1l}&=\frac{8\pi^2}{3}a^4\Big[-\frac{\left(V''(\Phi)\right)^2}{64\pi^2}\log N^2+\frac{12}{a^2}\frac{V''(\Phi)}{384\pi^2}\left(N^2+2\left(1-6\xi\right)\log N^2\right)\Big]\nonumber\\
	&+\frac{N^4}{48}\left(-1+2\log N^2\right)-\frac{N^2}{72}\left(13-72\xi+3\log N^2\right)+\Big(2\xi\left(1-3\xi\right)-\frac{29}{180}\Big)\log N^2\nonumber\\
	&+\mathcal H(a^2\,V''(\Phi))+\mathcal{O}\left(N^{-2}\right)\,,\label{deltaS}
\end{align}
where $\mathcal H(a^2\,V''(\Phi))$ contains only UV-finite ($N$-independent) terms. Its expression is given in the Appendix.

Up to now, we have not considered any specific form of $V(\Phi)$. Let us take the self-interacting potential {\small $V(\Phi)=\frac{m^2}{2}\Phi^2+\frac{\lambda}{4!}\Phi^4$}.
For the one-loop effective action {\small $\Gamma^{1l}=S^{(a)}[\Phi]+\delta S^{1l}$} we have (the inessential terms in the third line of\,\,\eqref{deltaS} are omitted)
{\begin{align}
		\Gamma^{1l}&=\frac{8\pi^2}{3}a^4\Big\{-\frac{1}{16\pi G}\Big[\,1-\frac{G\,m^2}{24\pi}\left(N^2+2(1-6\xi)\log N^2\right)\Big]\frac{12}{a^2}+\frac{\Lambda_{\rm cc}}{8\pi G}\Big[\,1-\frac{G\,m^4}{8\pi\Lambda_{\rm cc}}\log N^2\,\Big]\nonumber\\
		&+\frac{\xi}{2}\Big[1+\frac{\lambda}{384\pi^2\,\xi}\Big(N^2+2\left(1-6\xi\right)\log N^2\Big)\Big]\frac{12\,}{a^2}\Phi^2\nonumber
	\end{align}
	\begin{align}
		&+\frac{m^2}{2}\Big[1-\frac{\lambda\,}{32\pi^2}\log N^2\Big]\,\Phi^2+\frac{\lambda}{4!}\Big[1-\frac{3\lambda\,}{32\pi^2}\log N^2\Big]\Phi^4\,\Big\}\nonumber\\
		&+\frac{N^4}{48}\left(-1+2\log N^2\right)-\frac{N^2}{72}\left(13-72\xi+3\log N^2\right)+\Big(2\xi\left(1-3\xi\right)-\frac{29}{180}\Big)\log N^2\,.
		\label{1l-effac1}
\end{align}}

\noindent
Eq.\,\eqref{1l-effac1} is the central result of the present work, and we will comment on its consequences in the next section. Before doing that, we find it useful to proceed with the evaluation of $\delta S^{1l}$ following the second strategy mentioned above, namely proper-time regularization. We will then conveniently discuss both results together. 

Since the operator $(-\widetilde \square+12\xi+a^2V''(\Phi))$ in\,\eqref{scalarres} is dimensionless, to regularize its determinant we introduce the dimensionless proper-time  $\tau$, with numerical lower integration bound $1/N^2$ ($N\gg1$)
\begin{equation}\label{propertime}
	{\rm det}(-\widetilde \square+12\xi+a^2V''(\Phi))=e^{-\int_{1/N^2}^{+\infty}\frac{\dd{\tau}}{\tau}\,{\rm K}(\tau)}\,.
\end{equation}
The kernel ${\rm K}(\tau)$ is ($\lambda_n$ and $D_n$ are the eigenvalues and degeneracies reported in\,\eqref{eigenvalues})
\begin{equation}\label{kernel}
	{\rm K}(\tau)= \sum_{n=0}^{+\infty}D_n\, e^{-\tau\left(\lambda_n+12\xi+a^2V''(\Phi)\right)}\,.
\end{equation}
To calculate the determinant, we now insert\,\eqref{kernel} in\,\eqref{propertime}, perform the integration over $\tau$, and then sum over $n$ with the help of the EML formula. For the reader's convenience, we report the formula here
\begin{align}
	\sum_{n=n_i}^{n_f}f(n)=\int_{n_i}^{n_f}\dd[]x\,f(x)+\frac{f(n_f)+f(n_i)}{2}+\sum_{k=1}^{p}\frac{B_{2k}}{(2k)!}\left(f^{(2k-1)}(n_f)-f^{(2k-1)}(n_i)\right)+R_{2p}\,,
	\label{EML}
\end{align}
where $p$ is an integer, $B_m$ are the Bernoulli numbers and the rest $R_{2p}$ is given by
\begin{align}
	\small
	R_{2p}=-\frac{1}{(2p)!}\int_{n_i}^{n_f}\dd[]x\,f^{(2p)}(x)B_{2p}(x-[x]),
	\label{rest}
\end{align}
with $B_n(x)$ the Bernoulli polynomials, $[x]$ the integer part of $x$, and $f^{(i)}$ the $i$-th derivative of $f$ with respect to its
argument. 

Expanding the resulting expression of $\delta S^{1l}$ for $N\gg 1$, we finally get
\begin{align}
	\delta S^{1l}&=\frac{8\pi^2}{3}a^4\Big[-\frac{\left(V''(\Phi)\right)^2}{64\pi^2}\log N^2+\frac{12}{a^2}\frac{V''(\Phi)}{384\pi^2}\left(N^2+2\left(1-6\xi\right)\log N^2\right)\Big]\nonumber\\
	&-\frac{N^4}{24}-\frac{1-6\xi}{6}\,N^2+\Big(2\xi\left(1-3\xi\right)-\frac{29}{180}\Big)\log N^2\nonumber\\
	&+\mathcal Z(a^2\,V''(\Phi))+\mathcal{O}\left(N^{-2}\right)\,,
	\label{proper-time1}
\end{align}
where $\mathcal Z(a^2\,V''(\Phi))$ contains UV-finite terms (no dependence on $N$) and is similar to the term $\mathcal H(a^2\,V''(\Phi))$ in\,\eqref{deltaS}. For later purposes, we observe that $\mathcal Z(a^2\,V''(\Phi))$ contains terms of the kind$\sim\log a^2$ that, when considering the relation\,\eqref{offshell} between $N$ and $\Lambda$, cancel similar logarithms that arise from the first two lines in\,\eqref{proper-time1}.

Considering as before the potential {\small $V(\Phi)=\frac{m^2}{2}\Phi^2+\frac{\lambda}{4!}\Phi^4$}, the one-loop effective action $\Gamma^{1l}$ becomes  (the inessential terms in the third line of\,\eqref{proper-time1} are omitted)

{\begin{align}
		\Gamma^{1l}&=\frac{8\pi^2}{3}a^4\Big\{-\frac{1}{16\pi G}\Big[\,1-\frac{G\,m^2}{24\pi}\left(N^2+2(1-6\xi)\log N^2\right)\Big]\frac{12}{a^2}+\frac{\Lambda_{\rm cc}}{8\pi G}\Big[\,1-\frac{G\,m^4}{8\pi\Lambda_{\rm cc}}\log N^2\,\Big]\nonumber\\
		&+\frac{\xi}{2}\Big[1+\frac{\lambda}{384\pi^2\,\xi}\Big(N^2+2\left(1-6\xi\right)\log N^2\Big)\Big]\frac{12\,}{a^2}\Phi^2\nonumber\\
		&+\frac{m^2}{2}\Big[1-\frac{\lambda\,}{32\pi^2}\log N^2\Big]\,\Phi^2+\frac{\lambda}{4!}\Big[1-\frac{3\lambda\,}{32\pi^2}\log N^2\Big]\Phi^4\,\Big\}\nonumber\\
		&-\frac{N^4}{24}-\frac{1-6\xi}{6}\,N^2+\Big(2\xi\left(1-3\xi\right)-\frac{29}{180}\Big)\log N^2\,.
		\label{1l-effac-pt}
\end{align}}

\noindent
Apart from $\Phi$ independent terms (fourth line of both equations), the two results\,\eqref{1l-effac1} and\,\eqref{1l-effac-pt} for $\Gamma^{1l}$ coincide. Therefore, to determine the radiative correction $\delta m^2$ to the mass in the following we can equivalently consider either of these two expressions. For definiteness, we will refer to\,\eqref{1l-effac-pt}.

In the next section,  we investigate the two possible relations\,\eqref{offshell} and\,\eqref{onshell} between $N$ and the UV cutoff $\Lambda$. We write the expression for the one-loop effective action $\Gamma^{1l}$ in both cases, from which we read the corresponding one-loop corrected expressions of the parameters $m^2$, $\lambda$ and $\xi$. In particular, we focus on the main concern of the present work, namely the correction $\delta m^2$ to $m^2$.

\section{One-loop corrected parameters}
\label{1lparameters}

Let us consider the constant background $\phi=\Phi$ and the potential $V(\Phi)=\frac{m^2}{2}\Phi^2+\frac{\lambda}{4!}\Phi^4$. Using in\,\eqref{1l-effac-pt} the relation\,\eqref{offshell}, namely $N=\Lambda\,a$, the one-loop effective action $\Gamma^{1l}$ becomes
\begin{align}
	\text{\small$\Gamma^{1l}$}&\text{\small$=\frac{8\pi^2}{3}a^4\Big\{-\frac{1}{16\pi G}\Big[\,1+\frac{1-6\xi}{12\pi}G\Lambda^2-\frac{G\,m^2}{24\pi}\left(2(1-6\xi)\log\frac{\Lambda^2}{m^2}\right)\Big]\frac{12}{a^2}+\frac{\Lambda_{\rm cc}}{8\pi G}\Big[\,1-\frac{G}{8\pi \Lambda_{\rm cc}}\Lambda^4$}\nonumber\\
	&\text{\small$+\frac{m^2 G}{\,4\pi\Lambda_{\rm cc}}\Lambda^2-\frac{G\,m^4}{8\pi\Lambda_{\rm cc}}\log\frac{\Lambda^2}{m^2}\,\Big]+\frac{\xi}{2}\Big[1+\frac{\lambda}{384\pi^2\,\xi}\Big(2\left(1-6\xi\right)\log\frac{\Lambda^2}{m^2}\Big)\Big]\frac{12\,}{a^2}\Phi^2$}\nonumber\\
	&\text{\small$+\frac{m^2}{2}\Big[1+\frac{\lambda\,\Lambda^2}{32\pi^2\,m^2}-\frac{\lambda\,}{32\pi^2}\log\frac{\Lambda^2}{m^2}\Big]\,\Phi^2+\frac{\lambda}{4!}\Big[1-\frac{3\lambda\,}{32\pi^2}\log\frac{\Lambda^2}{m^2}\Big]\Phi^4\,\Big\}$}\nonumber\\
	&\text{\small$+\Big(2\xi\left(1-3\xi\right)-\frac{29}{180}\Big)\log\frac{\Lambda^2}{m^2}$}\,.
	\label{1ltrS2}
\end{align} 
Comparing $\Gamma^{1l}$ in\,\,\eqref{1ltrS2} with the classical (bare) action $S^{(a)}$ in\,\eqref{scalaraction2}, we see that $\Gamma^{1l}$ depends on $a$ and $\Phi$ in the same way as $S^{(a)}[\Phi]$. We can then easily read the radiative corrections to $m^2$, $\lambda$ and $\xi$, and for the one-loop corrected parameters {\small $m^2_{1l}$}, {\small $\lambda^{1l}$} and $\xi^{1l}$ we find
\begin{align}
	m^2_{1l}&=m^2\Big[1+\frac{\lambda\,\Lambda^2}{32\pi^2\,m^2}-\frac{\lambda\,}{32\pi^2}\log\frac{\Lambda^2}{m^2}\Big]\label{1l-m*}\\
	\lambda^{1l}&=\lambda\Big[1-\frac{3\lambda\,}{32\pi^2}\log \frac{\Lambda^2}{m^2}\Big]\label{1l-lambda*}\\
	\xi^{1l}&=\xi\Big[1+\frac{\lambda}{384\pi^2\,\xi}\Big(2\left(1-6\xi\right)\log\frac{\Lambda^2}{m^2}\Big)\Big]\,.\label{1l-xi*}
\end{align}
As anticipated, if the transition between the numerical cut $N$ and the UV cutoff $\Lambda$ is realized through the background radius $a$ (off-shell relation), namely resorting to\,\eqref{offshell}, the radiative correction to the mass presents the usual quadratic sensitivity to the UV scale $\Lambda$ (see\,\eqref{1l-m*}). Moreover, we see that the other two parameters $\lambda$ and $\xi$ possess the usual logarithmic sensitivity to $\Lambda$.

Let us consider now the other possible relation between $N$ and $\Lambda$, namely the on-shell relation\,\eqref{onshell}. In this case, for $\Gamma^{1l}$ we have
{\begin{align}
		\Gamma^{1l}&=\frac{8\pi^2}{3}a^4\Big\{-\frac{1}{16\pi G}\Big[\,1-\frac{G\,m^2}{24\pi}\Big( a_{\rm m}^{\,2} \Lambda^2+2(1-6\xi)\log (a_{\rm m}^{\,2}\Lambda^2)\Big)\Big]\frac{12}{a^2}+\frac{\Lambda_{\rm cc}}{8\pi G}\Big[\,1-\frac{G\,m^4}{8\pi\Lambda_{\rm cc}}\log (a_{\rm m}^{\,2}\Lambda^2)\,\Big]\nonumber\\
		&+\frac{\xi}{2}\Big[1+\frac{\lambda}{384\pi^2\,\xi}\Big(a_{\rm m}^{\,2}\Lambda^2+2\left(1-6\xi\right)\log (a_{\rm m}^{\,2}\Lambda^2)\Big)\Big]\frac{12\,}{a^2}\Phi^2\nonumber\\
		&+\frac{m^2}{2}\Big[1-\frac{\lambda\,}{32\pi^2}\log (a_{\rm m}^{\,2}\Lambda^2)\Big]\,\Phi^2+\frac{\lambda}{4!}\Big[1-\frac{3\lambda\,}{32\pi^2}\log (a_{\rm m}^{\,2}\Lambda^2)\Big]\Phi^4\,\Big\}\nonumber\\
		&-\frac{1}{24}\,a_{\rm m}^4\Lambda^4-\frac{1-6\xi}{6}\,a_{\rm m}^{\,2}\Lambda^2+\Big(2\xi\left(1-3\xi\right)-\frac{29}{180}\Big)\log (a_{\rm m}^{\,2}\Lambda^2)\,.
		\label{1l-effac2pt}
\end{align}}

\noindent
Here we recall that $a_{\rm m}$ is the radius that minimizes\,\footnote{The minimum $a_{\rm m}$ of $S^{(a)}[\Phi]$ is obtained solving the classical equations of motion for $a$ and $\Phi$, and depends on the parameters in $S^{(a)}[\Phi]$ and on the minimum $\Phi_{\rm m}$. For $m^2>0$ and $\xi>0$, \,$a_{\rm m}$\, is the de Sitter solution $a_{\rm m}=\sqrt{3/\Lambda_{\rm cc}}$ \,(with $\Phi_{\rm m}=0$)\,.} the action $S^{(a)}[\Phi]$ in\,\eqref{scalaraction2}. For the one-loop corrected parameters, from\,\eqref{1l-effac2pt} we get
\begin{align}
	m^2_{1l}&=m^2\Big[1-\frac{\lambda\,}{32\pi^2}\log (a_{\rm m}^{\,2}\Lambda^2)\Big]\label{1l-m.}\\
	\lambda^{1l}&=\lambda\Big[1-\frac{3\lambda\,}{32\pi^2}\log (a_{\rm m}^{\,2}\Lambda^2)\Big]\label{1l-lambda.}\\
	\xi^{1l}&=\xi\Big[1+\frac{\lambda}{384\pi^2\,\xi}\Big(a_{\rm m}^{\,2}\Lambda^2+2\left(1-6\xi\right)\log (a_{\rm m}^{\,2}\Lambda^2)\Big)\Big]\,.\label{1l-xi.}
\end{align}

Few comments are in order. From\,\eqref{1l-lambda.} we see that the quartic self-coupling $\lambda$ receives only a logarithmic correction. This is the usual result.
On the contrary, it is immediately apparent that the result for $m_{1l}^2$ in\,\eqref{1l-m.} is significantly different from the usual one: {\it no quadratic divergence} appears in the one-loop correction $\delta m^2$. Actually, $\delta m^2$ goes like $\log\Lambda$ rather than $\Lambda^2$. In this respect, we stress that the usual result $\delta m^2 \sim \Lambda^2$ enforces a $\Lambda^2$ dependence in the bare mass $m^2 (\Lambda)$, with a coefficient that must be finely tuned for it to cancel (almost exactly) this quadratically sensitive contribution. Differently, Eq.\,\eqref{1l-m.} shows that $m^2_{1l}\sim m^2$, so that the bare mass $m^2(\Lambda)$ may well be $m^2(\Lambda)\ll\Lambda^2$.  

Let us move to the one-loop corrected non-minimal coupling $\xi^{1l}$, Eq.\,\eqref{1l-xi.}. We see that, in addition to a logarithmic correction (the one usually found), $\xi$ receives a quadratically divergent contribution. The comparison of\,\eqref{1l-m.}-\,\eqref{1l-xi.} with the usual results\,\eqref{1l-m*}-\,\eqref{1l-xi*} shows that, when resorting to the on-shell relation\,\eqref{onshell} between $N$ and $\Lambda$, the UV behaviours of $m^2$ and $\xi$ are inverted: $m^2$ turns out to be only logarithmically sensitive to the  physical cutoff $\Lambda$, while $\xi$ seems to carry a quadratic sensitivity. Though up to now we did not have any reason to prefer one of the two relations, this latter result might raise some doubts on the on-shell relation\,\eqref{onshell}. In fact, being the non-minimal coupling $\xi$ a dimensionless parameter, a quadratic sensitivity to $\Lambda$ appears to be suspicious.

Finally, we would like to conclude with a remark concerning the \vv flat limit'' $a\to\infty$, where we expect to recover the usual flat space results. In such a limit, the effective potential $V^{1l}\equiv\Gamma^{1l}/\big(8\pi^2 a^4/3\big)$ ($\frac{8\pi^2}{3}a^4$ is the volume) turns out to be
{\footnotesize\begin{align}
	V^{1l}_{a\to\infty}&=\frac{\Lambda_{\rm cc}}{8\pi G}\Big[\,1-\frac{G}{8\pi \Lambda_{\rm cc}}\Lambda^4+\frac{m^2 G}{\,4\pi\Lambda_{\rm cc}}\Lambda^2-\frac{G\,m^4}{8\pi\Lambda_{\rm cc}}\log\frac{\Lambda^2}{m^2}\,\Big]\nonumber\\
	&+\frac{m^2}{2}\Big[1+\frac{\lambda\,\Lambda^2}{32\pi^2\,m^2}-\frac{\lambda\,}{32\pi^2}\log\frac{\Lambda^2}{m^2}\Big]\,\Phi^2+\frac{\lambda}{4!}\Big[1-\frac{3\lambda\,}{32\pi^2}\log\frac{\Lambda^2}{m^2}\Big]\Phi^4\,,
\end{align}}
if we resort to the off-shell relation\,\eqref{offshell}, and
{\footnotesize\begin{equation}
	V^{1l}_{a\to\infty}=\frac{\Lambda_{\rm cc}}{8\pi G}\Big(\,1-\frac{G\,m^4}{8\pi\Lambda_{\rm cc}}\log (a_{\rm m}^{\,2}\Lambda^2)\,\Big)+\frac{m^2}{2}\Big(1-\frac{\lambda\,}{32\pi^2}\log (a_{\rm m}^{\,2}\Lambda^2)\Big)\,\Phi^2+\frac{\lambda}{4!}\left(1-\frac{3\lambda\,}{32\pi^2}\log (a_{\rm m}^{\,2}\Lambda^2)\right)\Phi^4\,,
\end{equation}}

\noindent
if we resort to the on-shell relation\,\eqref{onshell}. From the above equations, we immediately see that only the off-shell relation\,\eqref{offshell} reproduces the flat space result for $V^{1l}$. This seems to be a further indication that the on-shell relation\,\eqref{onshell} might be problematic.

\section{Conclusions}
\label{conclusions}

In this work we have revisited the physical cutoff problem for the mass of a scalar field in the presence of gravity. More specifically, we have calculated the one-loop effective action of a self-interacting scalar field, non-minimally coupled to gravity, on a spherical background of radius $a$. Particular attention has been paid to the definition of the functional integral. Following Fradkin and Vilkovisky, we have used the diffeomorphism invariant path-integral measure obtained from the Hamiltonian formulation of the theory. An important consequence of this measure is that the fluctuation operator entering the one-loop determinant is automatically dimensionless. Its determinant can therefore be regularized by introducing a numerical cut $N$ on the number of eigenvalues of the dimensionless Laplace-Beltrami operator.

We have evaluated the one-loop effective action in two different ways: by directly truncating the sum over eigenvalues and by using proper-time regularization with a dimensionless lower cutoff $1/N^2$. Apart from field-independent terms, the two procedures give the same UV-dependent contributions to the effective action and, in particular, the same radiative corrections to the scalar mass $m^2$, the quartic coupling $\lambda$, and the non-minimal coupling $\xi$. This agreement shows that the results relevant for the present analysis do not depend on the specific implementation of the spectral regularization.

The crucial issue is then how the numerical cut $N$ has to be related to the UV cutoff $\Lambda$. We have considered two possibilities. If the relation is defined in terms of the off-shell background radius ($N=\Lambda \,a\,$), the standard result is recovered: the scalar mass receives a quadratically UV-sensitive contribution, $\delta m^2\sim\lambda\Lambda^2$, whereas the quartic coupling and the non-minimal coupling display their usual logarithmic dependence on $\Lambda$. Thus, within this prescription, the embedding of the scalar theory in a curved gravitational background does not modify the conventional physical cutoff problem.

A qualitatively different result is obtained if $N$ is related to $\Lambda$ through the on-shell radius $a_{\rm m}$ that extremizes the classical action ($N=\Lambda\, a_{\rm m}$). In this case the quadratic contribution to the scalar mass disappears and the UV sensitivity becomes only logarithmic, $\delta m^2\sim \lambda m^2\log\Lambda$\,.
Consequently, the bare mass is no longer forced by radiative corrections to be of order $\Lambda^2$. Taken by itself, this result would suggest that the presence of a non-trivial gravitational background may substantially soften the UV sensitivity of a scalar mass.

There are, however, important indications that prevent such a conclusion from being drawn at this stage. First, with the same on-shell identification the non-minimal coupling $\xi$, although dimensionless, acquires a contribution proportional to $a_{\rm m}^2\Lambda^2$. In this sense, the quadratic UV sensitivity is not simply removed but it is transferred from the scalar mass to the curvature coupling. More importantly, we have shown that the on-shell prescription does not reproduce the standard flat-space effective potential in the limit $a\to\infty$, while the off-shell relation $N=\Lambda \,a$ does. The flat-space limit therefore provides a consistency test that appears to favour the off-shell identification of the physical cutoff.

Our analysis thus highlights a point that is more general than the specific model considered here. Once the functional determinant is naturally expressed in terms of a dimensionless numerical cut, the UV behaviour assigned to the parameters of the theory depends crucially on the physical prescription used to convert this numerical cut into an energy scale. This identification cannot be regarded as a merely technical step: it is part of the physical definition of the cutoff and must satisfy appropriate consistency requirements. In the present setting, the requirement of recovering the flat-space theory singles out the off-shell relation and, correspondingly, restores the usual quadratic sensitivity of the scalar mass.

The results obtained here therefore do not provide, by themselves, a gravitational solution to the physical cutoff problem of the Higgs mass. Rather, they show that gravity opens a new and non-trivial question concerning the physical identification of the UV scale in a curved spacetime. Determining whether alternative, physically consistent prescriptions can soften the UV sensitivity of scalar masses requires extending the present analysis beyond the spherical background and, ultimately, including the full set of Standard Model and gravitational fluctuations. It would also be interesting to investigate whether the conclusions reached here persist on more general backgrounds and how they are affected when the geometry itself is treated quantum mechanically.

\section*{Acknowledgments}
The work of CB has been supported by the European Union – Next Generation EU through the research grant number P2022Z4P4B “SOPHYA - Sustainable Optimised PHYsics Algorithms: fundamental physics to build an advanced society” under the program PRIN 2022 PNRR of the Italian Ministero dell’Università e Ricerca (MUR). The work of VB, FC, RG and AP is carried out within the INFN project  QGSKY.

\section*{Appendix}

In this Appendix we report the function $\mathcal{H}(a^2 V''(\Phi))$ that appears in Eq.\,\eqref{deltaS}. As already stressed  in the text, its specific form is not of interest for the analysis presented in this work, where we are focused on the UV-sensitivity of the one-loop effective action $\Gamma^{1l}$. For completeness, however, we write its expression below
\begin{align}
	&\text{\small$\mathcal{H}(a^2 V'')=\frac{1}{12} \Big(-\sqrt{9-4 a^2 V''-48 \xi} \Big(a^2 V''+12 \xi -2\Big) \log \Gamma\Big(\frac{3+\sqrt{9-4 a^2 V'' -48 \xi}}{2} \Big)$}\nonumber\\
	&\text{\small$+\Big(a^2 V''+12 \xi -2\Big) \sqrt{9-4 a^2 V''-48 \xi} \,\log \Gamma\Big(\frac{3-\sqrt{9-4 a^2 V'' -48 \xi}}{2} \Big)$}\nonumber\\
	&\text{\small$-12 \Big[\psi ^{(-4)}\Big(\frac{3+\sqrt{9-4 a^2 V'' -48 \xi}}{2} \Big)+\psi ^{(-4)}\Big(\frac{3-\sqrt{9-4 a^2 V'' -48 \xi}}{2}\Big)\Big]$}\nonumber\\
	&\text{\small$+\Big(6 a^2 V''+72 \xi -13\Big) \Big[\psi ^{(-2)}\Big(\frac{3+\sqrt{9-4 a^2 V'' -48 \xi}}{2}\Big)+\psi ^{(-2)}\Big(\frac{3-\sqrt{9-4 a^2 V'' -48 \xi}}{2}\Big)\Big]$}\nonumber\\
	&\text{\small$+6 \sqrt{9-4 a^2 V''-48 \xi} \Big[\psi ^{(-3)}\Big(\frac{3+\sqrt{9-4 a^2 V'' -48 \xi}}{2}\Big)-\psi ^{(-3)}\Big(\frac{3-\sqrt{9-4 a^2 V'' -48 \xi}}{2}\Big)\Big]$}\nonumber\\
	&\text{\small$-4 \zeta '(-3)\Big)-\frac{a^2 V''}{12}+\frac{13 \log (A)}{6}-\xi +\frac{3 \zeta (3)}{8 \pi ^2}+\frac{319}{2160}+\frac{1}{2} \log (2 \pi )$}\,.
\end{align}
In the above equation, $A$ is the Glaisher's constant ($A\simeq1.282427$), $\zeta(z)$ is the Riemann zeta function ($\zeta (3)\simeq1.20206$ and $\zeta '(-3)\simeq0.00538$), and $\psi ^{(-n)}(z)$ (with $n$ positive integer) are the polygamma functions of negative order defined as\,\cite{Adamchik}
\begin{equation}
	\psi ^{(-n)}\left(z\right)=\frac{1}{(n-2)!}\int_{0}^z\dd{t}(z-t)^{n-2} \log\,\Gamma\left(t\right)\qquad\qquad \text{for}\,\,\Re(z)>0\,.
\end{equation}

\end{document}